\documentclass[prd,aps,preprint,amsmath,nofootinbib,amssymb,eqsecnum,showkeys]{revtex4-1}

\usepackage{subfigure}
\usepackage{graphicx}
\usepackage{xcolor}
\newcommand{\be}{\begin{equation}}
\newcommand{\ee}{\end{equation}}
\newcommand{\bea}{\begin{eqnarray}}
\newcommand{\eea}{\end{eqnarray}}

\begin{document}
\title{Leptogenesis in fast expanding Universe}

\author{Shao-Long Chen}
\email[E-mail: ]{chensl@mail.ccnu.edu.cn}
\affiliation{Key Laboratory of Quark and Lepton Physics (MoE) and Institute of Particle Physics, Central China Normal University, Wuhan 430079, China}

\author{Amit Dutta Banik}
\email[E-mail: ]{amitdbanik@mail.ccnu.edu.cn}
\affiliation{Key Laboratory of Quark and Lepton Physics (MoE) and Institute of Particle Physics, Central China Normal University, Wuhan 430079, China}

\author{Ze-Kun Liu}
\email[E-mail: ]{zekunliu@mails.ccnu.edu.cn}
\affiliation{Key Laboratory of Quark and Lepton Physics (MoE) and Institute of Particle Physics, Central China Normal University, Wuhan 430079, China}

\date{\today}

\begin{abstract}
With the consideration of a fast expanding Universe in effect due to an additional scalar field, 
we present a study of leptogenesis in non-standard cosmology. 
The Hubble expansion rate is modified by the new added scalar field $\varphi$, which can change the abundance of lepton asymmetry 
resulted by the leptogenesis mechanism. We report a significant deviation from the standard 
unflavored leptogenesis scenario can be achieved in presence of the scalar field $\varphi$  that 
dominates the energy budget of the early Universe. We present our results for leptogenesis from type-I
seesaw with heavy right-handed Majorana neutrinos. The results are based on Boltzmann equations and effects of the scalar field are  
similar for other kinds of leptogenesis framework.
\end{abstract}


\maketitle

\section {Introduction}
With the discovery of Higgs boson at large Hadron collider (LHC)~\cite{Aad:2012tfa,Chatrchyan:2012xdj}, the
Standard Model (SM) is undoubtedly established as the most successful theory of particle 
physics. However neutrino mass, matter-antimatter asymmetry in the Universe and the identity of dark matter
motivate the search for the new physics beyond the Standard Model (BSM). An elegant and economic way to explain 
neutrino mass and matter-antimatter asymmetry simultaneously in a single framework is to add heavy right-handed 
Majorana neutrinos into the Standard Model. The heavy Majorana neutrinos not only generate tiny 
neutrino mass through type-I seesaw mechanism~\cite{Minkowski:1977sc}-\cite{Schechter:1980gr}, 
but also provide the necessary source of CP asymmetry and lepton number violation. Once out of thermal 
equilibrium, decay of heavy right-handed neutrinos (RHNs) generate a lepton asymmetry. 
The lepton asymmetry generated is partially transferred into baryon asymmetry that explains matter-antimatter 
asymmetry in the Universe. The process is known as leptogenesis in common. 
Detailed studies of leptogenesis including the Boltzmann equations and dynamics of the process can be found in 
many literatures~\cite{Fukugita:1986hr}-\cite{Dolan:2018qpy}. It is also possible to obtain
matter-antimatter asymmetry by leptogenesis from different seesaw mechanisms by adding triplet scalar (or triplet fermion) 
known as type-II (type-III) seesaw~\cite{Mohapatra:1980yp}-\cite{Franco:2015pva}, as well as the radiative seesaw models
~\cite{Kashiwase:2012xd}-\cite{Mahanta:2019gfe}. 
In this study, we focus on the study of the Boltzmann equations for leptogenesis. For the purpose of 
demonstration we restrict our study of leptogenesis obtained from right-handed neutrinos in the 
type-I seesaw.

The evolution of the Universe in standard cosmology is specified into two era, radiation era and matter era. The era before Big Bang Nucleosynthesis (BBN) is 
known as radiation era while after BBN the Universe is matter dominated. Although we have observational evidences from the period after BBN which refers to 
the matter era, there is no signature that confirm the Universe is only radiation dominated before BBN. Therefore, there exists possibility that before BBN the 
cosmology of the Universe is different from purely radiation era. Such possibilities have been explored in literatures with various non-standard cosmological 
models such as including a new scalar field~\cite{DEramo:2017gpl}, late decay of inflation field~\cite{Arbey:2008kv,Arbey:2011gu}, quintessence models
~\cite{Salati:2002md}-\cite{Pallis:2005hm}, and anisotropic expansion of the Universe~\cite{Barrow:1982ei,Kamionkowski} etc. 
These alternate cosmology scenarios deviate from standard cosmology and predict different expansion rate of Universe, therefore,  the 
evolution of the Universe is changed in the framework of alternate cosmology. If this happens in the era of leptogenesis, 
the standard leptogenesis process will be modified and affect the abundance of lepton asymmetry from the decay of right-handed neutrinos. 
It was shown in~\cite{DEramo:2017gpl} that presence of a scalar field $\varphi$ can result in a fast 
expanding Universe, and therefore affects the thermal freeze-out relics of dark matter. Similar studies with the fast expanding 
Universe were performed for the case of freeze-in dark matter~\cite{DEramo:2017ecx,Maldonado:2019qmp}, keV neutrino dark matter~\cite{Biswas:2018iny,Fernandez:2018tfa}, asymmetric dark matter models~\cite{Iminniyaz:2013cla}-\cite{Iminniyaz:2018das}
and sterile neutrinos~\cite{Gelmini:2019esj,Gelmini:2019wfp}. 
Study of leptogenesis  was performed earlier with different cosmological scenarios of the Universe such as including dark matter ~\cite{Bernal:2017zvx}
or scalar tensor gravity \cite{ Dutta:2018zkg}.
In this work, under the framework of fast expanding Universe, we study the characteristics of 
leptogenesis in modified cosmology. We investigate the influence of a scalar $\varphi$ on Boltzmann equations (BEs) for leptogenesis and 
its effects on the evolution of lepton asymmetry. Leptogenesis in non-standard cosmology can therefore be referred as non-standard leptogenesis. 
 The paper is organized as follows. In Sec.~\ref{cosmo}, we briefly describe the fast expanding Universe. In the next section, we derive the Boltzmann equations for leptogenesis  
in the modified Universe. Study of non-standard leptogenesis in fast expanding Universe and its comparison with standard leptogenesis is presented in 
Sec~\ref{res}. Finally in Sec.~\ref{con}, the paper is summarized with concluding remarks.

\section{Non-standard Cosmology of modified Universe}
\label{cosmo}
In standard cosmology, before BBN the energy density in the Universe is assumed to be
 governed by the radiation. Energy density of radiation can be expressed as 
\be
\rho_{rad}=\frac{\pi^2}{30}g_*T^4
\ee
where $T$ denotes the temperature of the Universe and $g_*$ is the effective relativistic degrees of freedom. 
Therefore the expansion rate of Universe, i.e., Hubble parameter $H$ depending on the radiation energy density is of the form
\be
H=\sqrt{\frac{8\pi G \rho_{rad}}{3}}=1.66 \sqrt{g_*} \frac{T^2}{M_P}\,,
\label{Hub}
\ee
with the Planck mass $M_P=1.22\times 10^{19}$ GeV.
It is conventionally considered that the era of radiation domination starts after inflation below the reheating temperature $T_{\rm RH}$. 
The situation may alter in presence of additional fields and it may not always be the case that dynamics of Universe is governed only by radiation
 in the large range of temperature from $T_{\rm RH}$ to $T_{\rm BBN}$. Let us consider that a scalar field $\varphi$
is also present at the early Universe and its energy density depends on the scale factor $a$
\be
\rho_{\varphi}\sim a^{-(4+n)}, \hskip 5mm n>0\,\, .
\label{rhophi}
\ee 
The entropy density of Universe at any temperature $T$ is given as
\be
s=\frac{2\pi^2}{45}g_{*s}T^3\,\, ,
\label{entropy}
\ee
where $g_{*s}$ is the entropy degrees of freedom. We consider that total entropy $S=sa^3$ is constant in 
a co-moving frame, which indicates at any temperature $T$, $g_{*s}T^3a^3$ is unchanged. 
Now we define a temperature $T_r$ when energy density of $\varphi$ becomes equal to energy density of
 radiation, i.e., at $T=T_r$, $\rho_{\varphi}=\rho_{rad}$. Using the relation described in Eq.~(\ref{rhophi}), we can write down the energy density of $\varphi$ as
\be
\rho_{\varphi}(T)=\rho_{\varphi}(T_r)\bigg[\frac{g_*(T_r)}{g_*(T)}\bigg(\frac{g_{*s}(T)}{g_{*s}(T_r)}\bigg)^{(4+n)/3}\bigg(\frac{T}{T_r}\bigg)^n\bigg]\,\, .
\label{phirad1}
\ee
Therefore, total energy density is expressed as
\be
\rho_{Tot}=\rho_{rad}+\rho_{\varphi}=\rho_{rad}\bigg[1+\frac{g_*(T_r)}{g_*(T)}\bigg(\frac{g_{*s}(T)}{g_{*s}(T_r)}\bigg)^{(4+n)/3}\bigg(\frac{T}{T_r}\bigg)^n\bigg]\,\, .
\ee
Assuming both $g_{*}$ and $g_{*s}$ to be constant and independent of temperature \footnote{This assumption is valid for large $T$ which is also true for the era of leptogenesis.}, we can rewrite the total energy density as
\be
\rho_{Tot}=\rho_{rad}\bigg[1+\bigg(\frac{T}{T_r}\bigg)^n\bigg]
\ee 
Therefore, the Hubble parameter can be redefined as
\be
H^\prime=1.66 \sqrt{g_*} \frac{T^2}{M_P}\bigg[1+\bigg(\frac{T}{T_r}\bigg)^n\bigg]^{1/2}
=H\bigg[1+\bigg(\frac{T}{T_r}\bigg)^n\bigg]^{1/2}\,\, .
\label{Hnew}
\ee
From Eq.~(\ref{Hnew}), we can easily observe that for $T\gg T_r$, the Universe will be expanding faster with respect to the radiation dominated Universe. 
It is worth mentioning that the temperature  $T_r$ should not be too small otherwise it can modify the BBN constraints as pointed out in \cite{DEramo:2017gpl}.  For a certain value of $n$, BBN constraints provide a lower limit on $T_r$,
\be
T_r\geq (15.4)^{1/n}\hskip 5 pt {\rm MeV}\,\, .
\label{Tr}
\ee
However, since we are interested to observe the effects of $\varphi$ in the era of leptogenesis, 
we consider $T_r$ to be sufficiently large avoiding the conflict with BBN constraints.

\section{Boltzmann equation for RHN decay and lepton asymmetry in modified Universe}
\label{LEP}
In this section we study leptogenesis with RHNs in type-I seesaw in the non-standard cosmology framework.
We start with the simplest scenario with hierarchical RHN mass and leptogenesis is governed by the decay of the lightest RHN of mass $M_1$. 
Right-handed Majorana neutrinos interacts with Standard Model leptons and the interaction Lagrangian is given as
\be
{\cal L}=-Y_{ij}\bar{l_i}\tilde{\Phi} N_{j} - \frac{1}{2}M_j\bar{N^c}_jN_j + h.c.\,\, ,
\label{L}
\ee
where $l$ denote lepton doublets and $\Phi$ is the Standard Model Higgs doublet. 
The mass term of RHNs is the source of lepton number violation as we assigning lepton number to right-handed Majorana neutrinos. 
For simplicity we consider the mass matrix for heavy Majorana neutrinos to be diagonal and assume they are
hierarchical $M_3>M_2>M_1$.
The Yukawa coupling $Y_{ij}$ are in general complex and provide the source of CP violation.

The lepton asymmetry generated from CP violating decay of $N_1$ can be expressed by an CP asymmetry parameter $\varepsilon$ which is expressed as
\begin{align}
\varepsilon & = \frac{\sum_{\alpha}[\Gamma(N_1 \rightarrow l_{\alpha}+\Phi)-\Gamma(N_1 \rightarrow \bar{l}_{\alpha}+\Phi^{*})]}{\Gamma_1}
\\  
&=-\frac{3}{16\pi}\frac{1}{(Y^{\dagger}Y)_{11}}\sum_{j=2,3}{\rm Im}[(Y^{\dagger}Y)^2_{1j}]\frac{M_1}{M_j}\,\, . 
\label{CPasy}
\end{align}
where $\Gamma_1=\frac{M_1}{8\pi}(Y^{\dagger}Y)_{11}$ denote the total decay width of $N_1$.

The light active neutrino mass obtained from type-I seesaw mechanism is expressed as
\be
M_{\nu}=-m_{D}^TM^{-1}m_{D}\,\, ,
\label{numass}
\ee
where $M$ is the mass matrix for heavy right-handed neutrinos considered to be diagonal. 
Using the Casas-Ibarra (CI) parametrization~\cite{Casas:2001sr} for neutrino Yukawa couplings, an upper limit of the CP asymmetry $\varepsilon$ can be obtained of the form
\be
|\varepsilon|< \frac{3}{16\pi v^2}M_1m_{\nu}^{\rm max}\,\, ,
\label{limit}
\ee
where $m_{\nu}^{\rm max}$ is the largest mass of light neutrinos. Considering $m_{\nu}^{\rm max}=\sqrt{\Delta m_{31}^2}$ from neutrino oscillation data~\cite{Patrignani:2016xqp}, for a given value of $|\varepsilon|$ one can obtain the well known Davidson-Ibarra bound~\cite{Davidson:2002qv} for lightest RHN mass. With a typical value of $|\varepsilon|=10^{-6}$, we get $M_1\geq 10^{10}$ GeV.
Therefore, the type-I seesaw can generate light neutrino mass and also provide successful leptogenesis to explain the matter-antimatter asymmetry in the Universe. Now let us get back to the discussions on Boltzmann equation. Considering leptogenesis from decay of lightest RHN $N_1$, the standard Boltzmann equations govern the scaled co-moving number density of lightest RHN is given as  
\begin{eqnarray}
\label{eq:BE}
\frac{d Y_{N_1}}{dz}&=& -\frac{\Gamma_1}{H z} 
\frac{K_1(z)}{K_2(z)}\left(Y_{N_1}-Y_{N_1}^{\rm eq}\right)\, .
\label{BE1}
\end{eqnarray}
where $Y_{N_1}=n_{N_1}/s$, $s$ is the co-moving  entropy density, and variable $z=M_{1}/T$. The equilibrium rescaled number density of 
$N_{1}$ is $Y_{N_1}^{\rm eq}$=$\frac{45g}{4\pi^4}\frac{z^2K_2(z)}{g_{*s}}$, and $K_{1,2}$ are modified Bessel functions.

Now in presence of the additional scalar $\varphi$, the Hubble parameter is replaced by the 
expression in Eq.~(\ref{Hnew}), and the Boltzmann equation (BE) in Eq.~(\ref{BE1}) is then modified as
\begin{eqnarray}
\frac{d Y_{N_1}}{dz}&=& -\frac{\Gamma_1}{H z 
\bigg[1+\left(\frac{M_1}{T_rz}\right)^{n}\bigg]^{1/2}} 
\frac{K_1(z)}{K_2(z)}\left(Y_{N_1}-Y_{N_1}^{\rm eq}\right)\, .
\label{BE2}
\end{eqnarray}
We further define ${H_1}(T=M_1)=1.66g_*^{1/2}M_1^2/M_P=Hz^2$, the BE becomes
\begin{eqnarray}
\frac{d Y_{N_1}}{dz}&=& 
-z\frac{\Gamma_1}{H_1}\bigg[1+\left(\frac{M_1}{T_r z}\right)^{n}\bigg]^{
-1/2}
\frac{K_1(z)}{K_2(z)}\left(Y_{N_1}-Y_{N_1}^{\rm eq}\right)\,,
\label{BE3}
\end{eqnarray}
which depends on the new parameters $T_r,n$ and ${\Gamma_1}/{H_1}$ is the standard decay parameter in leptogenesis.

Similarly, one can derive the BE for lepton asymmetry in the modified 
framework, which is given as
\begin{eqnarray}
\frac{d Y_{L}}{dz} &=&  - \frac{\Gamma_1}{H_1} 
\bigg[1+\left(\frac{M_1}{T_rz}\right)^{n}\bigg]^{-1/2}
\left (\varepsilon  z \frac{K_1(z)}{K_2(z)}(Y_{N_1}^{eq}- Y_{N_1})  +  
\frac{z^{3} K_1(z)}{4} Y_L  \right)\, \, .
\label{asy}
\end{eqnarray}

It is easily observed that Eq.~(\ref{BE3}) and Eq.~(\ref{asy}) reduces 
to standard BEs for leptogenesis with Hubble parameter $H$ in absence of the scalar field $\varphi$, which is also valid for $T_r\gg M_1$. However, in that case, the Universe will be radiation dominated  when $T\sim M_1$ and we go back to the usual leptogenesis. Therefore, in order to observe the effect of modified evolution of the Universe, we consider a range of $T_r$ which is comparable with $M_1$.  It is to be noted that the Boltzmann equation for lepton asymmetry given in Eq.~(\ref{asy})
is valid in absence of flavor effect. Flavor effects can be important for the study of leptogenesis when charged lepton Yukawa interactions becomes fast as they interact with thermal bath. Comprehensive studies on flavor effects in leptogenesis can be found in literatures~\cite{Abada:2006fw}-\cite{Dev:2017trv}.
Further studies of flavor leptogenesis provides limit on RHN mass $M_1$ and it is found that for $M_1\geq5\times 10^{11}$ GeV, flavor effects can be neglected. In this work, we consider a conservative approach assuming $M_1\geq5\times 10^{11}$ GeV and hence unflavored approximation of Boltzmann equation is valid throughout the analysis\,\footnote{In a recent work \cite{Croon:2019dfw}, it is found that for minimal leptogenesis (type-I seesaw) the reheating temperature $T_{\rm RH}$ could be as large as $T_{\rm RH}\leq 1.1\times 10^{13}$ GeV which is in agreement with the choice of $M_1$. It is to be noted that in case of supersymmetric models, gravitino overproduction sets an upper bound on reheating temperature $T_{\rm RH}$ which contradicts with the choice of high RHN mass \cite{Khlopov:1984pf,Khlopov:1993ye}. This can be avoided if right-handed neutrinos are produced non-thermally or if one considers non-supersymmetric extensions \cite{Davidson:2002qv}. Alternatively, it is also possible to lower the scale of leptogenesis in supesymmetric scenarios \cite{Hamaguchi:2001gw,Grossman:2005yi}.}. Since, as the expansion rate of Universe is higher due to presence of $\varphi$, we expect charged lepton Yukawa interactions to enter in thermal bath later at some lesser temperature. This indeed validates our conservative choice of RHN mass. The amount of lepton asymmetry produced is transferred into baryon asymmetry via Spahleron transition process. Assuming Sphaleron is active before electroweak phase transition, the relation between net baryon asymmetry generated $Y_{B}$ with $Y_L$ is expressed as~\cite{Davidson:2008bu}
\be  
Y_B=\frac{8n_f+4n_{\phi}}{22n_f+13n_{\phi}}Y_L\,\, ,
\label{relation}
\ee
where $n_f=3$ is the number of fermion generations and $n_{\phi}=1$ as we have only one scalar field. Therefore, Eq.~(\ref{relation}) reduces to
\be
Y_B=\frac{28}{79}Y_L\,\\, .
\label{BL}
\ee
The abundance of Baryon asymmetry in the Universe as obtained from Planck measurements is $Y_{B}=(8.24-9.38)\times 10^{-11}$ \cite{Patrignani:2016xqp}. 
Therefore, the amount of lepton asymmetry required in order to produce observed Baryon asymmetry is $Y_L=(2.37-2.70)\times10^{-10}$.

\section{Non-standard leptogenesis: Observations and Results}
\label{res}
In this section we present the effects of modified cosmology due to scalar $\varphi$
on the standard leptogenesis. The input parameters for the Boltzmann equations Eqs.~(\ref{BE3})-(\ref{asy}) are,
\be
{\Gamma_1}/{H_1}=K,\hskip 2mm \varepsilon, \hskip 2mm T_r/M_1,\hskip 2mm n\,\, .
\label{parameters}
\ee
We consider the case that $T_r/M_1\leq1$ to observe the effects of $\varphi$ on leptogenesis. The factor $K=\frac{\Gamma_1}{H_1}$, is the well known decay parameter which determines the washout effects of asymmetry. Boltzmann equations expressed in Eqs.~(\ref{BE3})-(\ref{asy}) also depends on initial values of  $Y_{N_1}$ and $Y_L$. We consider there is no initial lepton asymmetry in the Universe, i.e., $Y_L^{\rm in}=0$. For RHN, we study cases with two initial conditions I) $Y_{N_1}^{in}=Y_{N_1}^{eq}$ and II)  $Y_{N_1}^{in}=0$. In this section, we will investigate for both the initial conditions of co-moving density of $N_1$.

\begin{figure}[!ht]
\centering
\subfigure[]{
\includegraphics[height=6 cm, width=7 cm,angle=0]{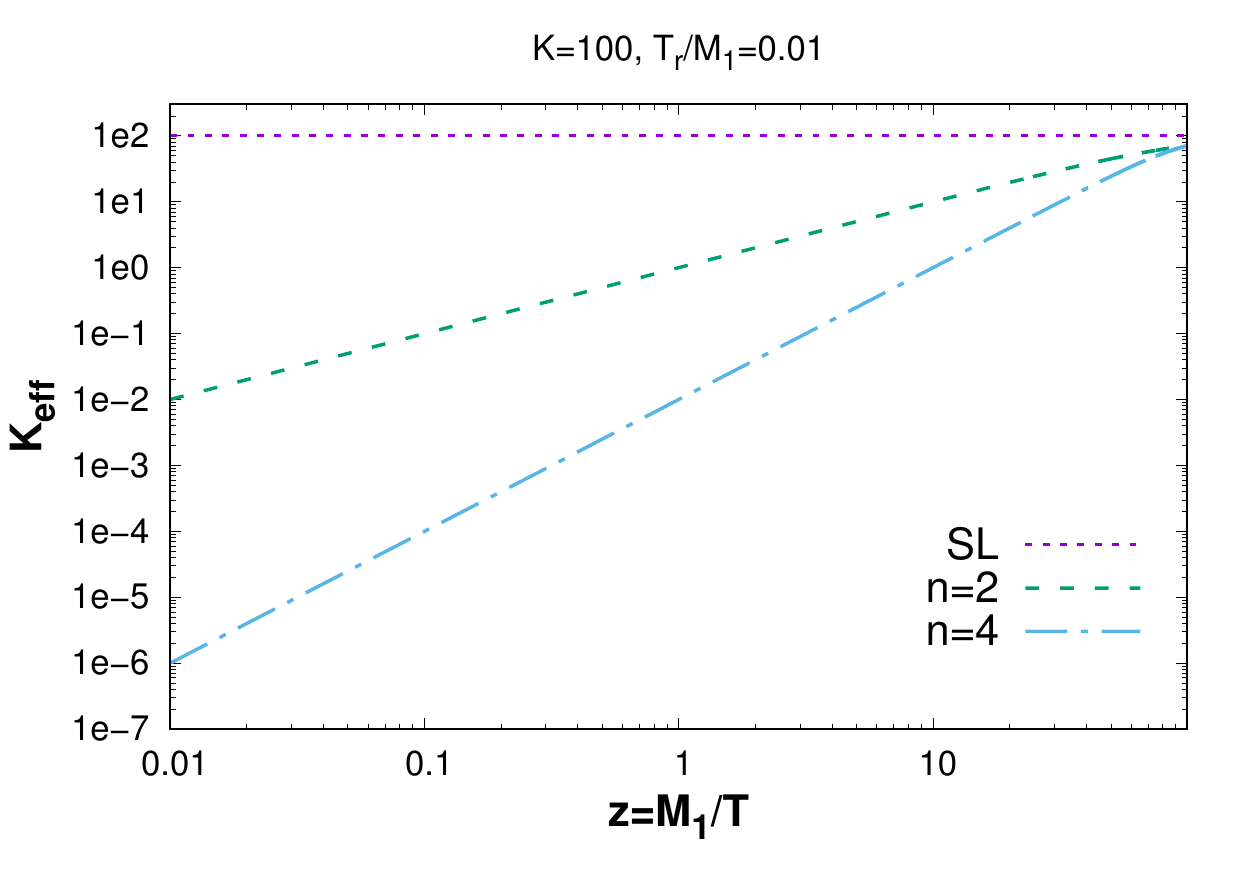}}
\subfigure []{
\includegraphics[height=6 cm, width=7 cm,angle=0]{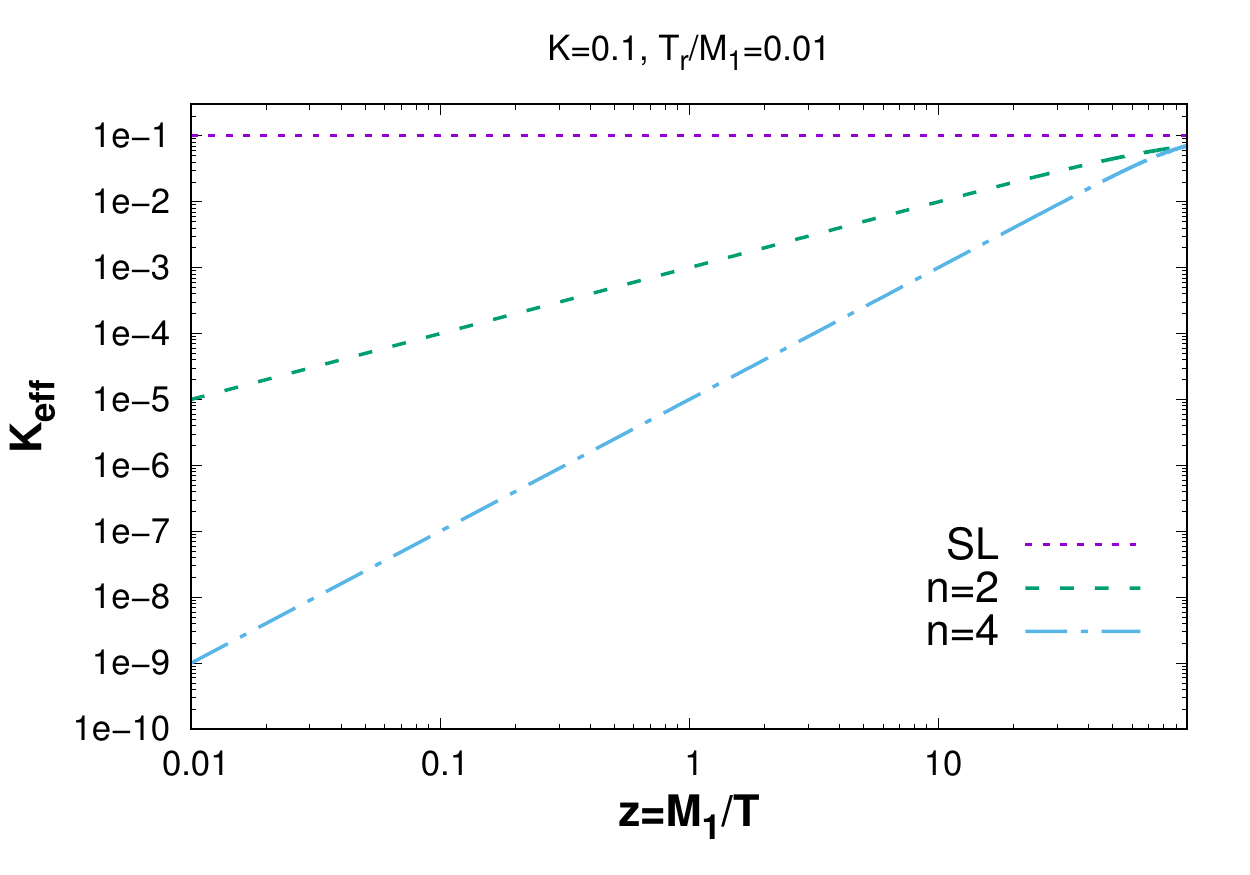}}
\caption{Variation of $K$ factor in non-standard leptogenesis ($K_{\rm eff}$) and comparison with standard leptogenesis (SL) for $K=100$ (left panel) and $K=0.1$ (right panel).}
\label{1}
\end{figure}

As we have  mentioned, the decay parameter $K$ plays an important role in leptogenesis and controls the washout of asymmetry generated in $Y_L$. Depending on the value of $K$, one can classify the regions of different washout scenarios defined as a) strong washout for $K>1$ and b) weak washout when $K<1$. In this work we will consider both the washout regimes with decay parameter having value $K=100$ for strong washout and $K=0.1$ for weak washout region. Looking into the expressions of Boltzmann equations Eqs.~(\ref{BE3})-(\ref{asy}), one can notice that an extra term $\bigg[1+\left(\frac{M_1}{T_rz}\right)^{n}\bigg]^{-1/2}$ appears in these equations along with the decay parameter $K$. Taking this into account we define a modified decay parameter $K_{\rm eff}=f(K,T_r/M_1,n,z)$ which is expressed as 
\be
K_{\rm eff}=K\bigg[1+\left(\frac{M_1}{T_rz}\right)^{n}\bigg]^{-1/2}\,\, .
\label{Knew}
\ee
Therefore, with this newly defined decay parameter $K_{\rm eff}$, Boltzmann equations becomes exactly identical to its original form as in standard leptogensis. Hence, we treat $K_{\rm eff}$
as the new decay parameter for leptogenesis in modified cosmology. In Fig. \ref{1}(a)-(b) we plot
the variation of effective decay parameter $K_{\rm eff}$ versus $z$ with a chosen value of $T_r/M=0.01$ for two values of $n=2,4$. Fig.~\ref{1}(a) is plotted for $K=100$ and the same with $K=0.1$ is shown in Fig.~\ref{1}(b).  Comparing the values of $K_{\rm eff}$ for $n=2$ with the standard leptogenesis where $K$ remains constant, in Fig.~\ref{1}(a) it can be easily observed that the decay parameters remains in the weak washout regime $K_{\rm eff}$ for smaller values of $z$ and enters into strong washout regime later at a larger value of $z\geq1$. Similar behaviour of $K_{\rm eff}$ is achieved for $n=4$, but in this case $K_{\rm eff}=1$ happens at value of $z\sim10$. This indicates that, for larger values of $n$, in non-standard leptogenesis, washout effect is suppressed considerably. A similar plot for $K_{\rm eff}$ is shown in Fig.~\ref{1}(b) for $K=0.1$ keeping other parameters fixed. Fig.~\ref{1}(b), we observe the same pattern as in Fig.~\ref{1}(a) and weak washout becomes much weaker in the fast expanding Universe with respect to the ordinary leptogenesis scenario. 
This is an obvious consequence of fast expanding Universe since as it expands faster, interactions fails to remain
in equilibrium. We will now discuss how this situation actually effects the evolution of lepton asymmetry. 

\subsection{Case I : $Y_{N_1}^{in}=Y_{N_1}^{eq}$, $\varepsilon=10^{-6}$}
\label{yeq}
In this section, we describe the evolution of co-moving number density of RHN $Y_{N_1}$ and lepton asymmetry $Y_L$ depending on parameters from non-standard cosmology $n,~T_r/M_1$. We assume that initially $N_1$ has equilibrium number density $Y_{N_1}^{in}=Y_{N_1}^{eq}$ and solve the Boltzmann equations Eq.~(\ref{BE3})-(\ref{asy}) for $\varepsilon=10^{-6}$ with two values of decay parameters $K=100,0.1$.  

\subsubsection*{Effects of $n$ on BE solutions}

\begin{figure}[!ht]
\centering
\subfigure[]{
\includegraphics[height=6 cm, width=7 cm,angle=0]{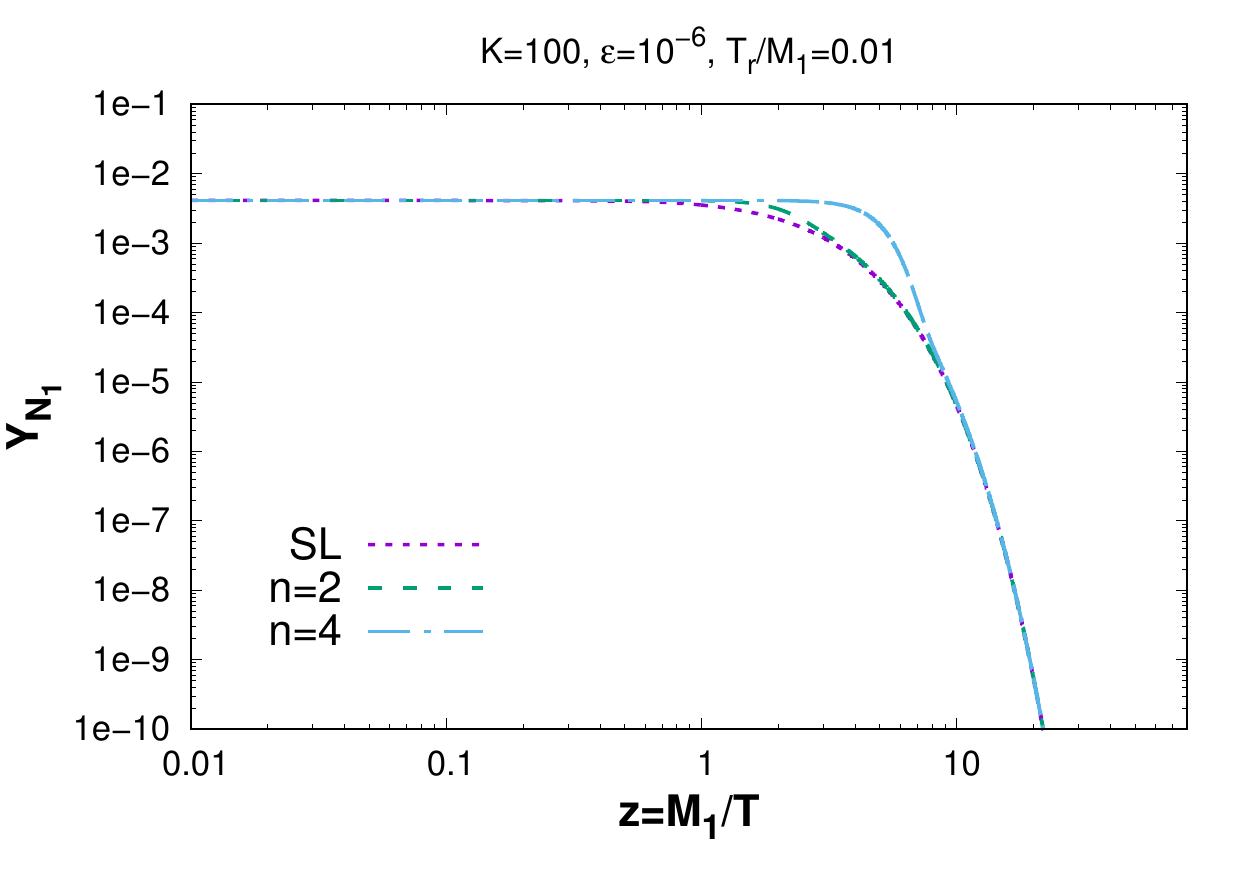}}
\subfigure []{
\includegraphics[height=6 cm, width=7 cm,angle=0]{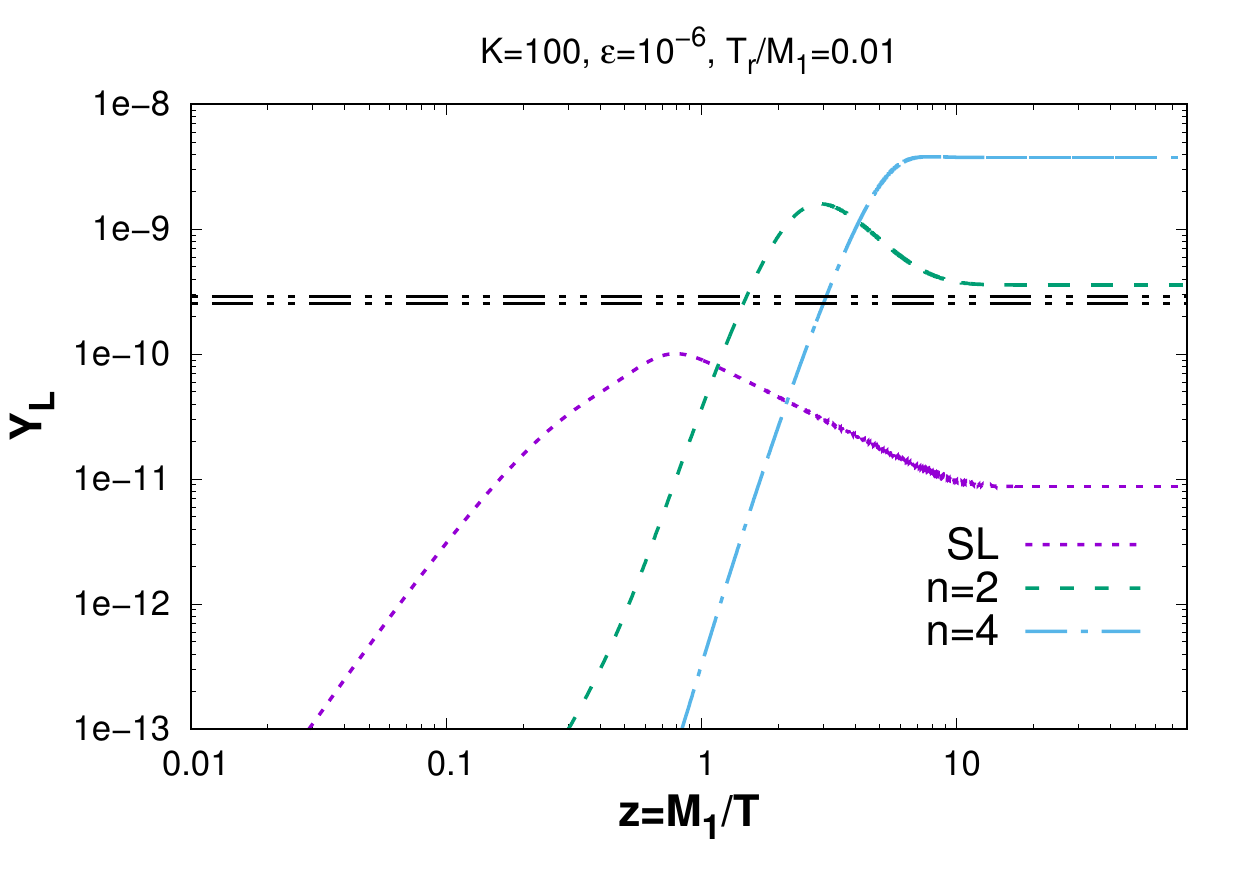}}
\caption{Variation of $Y_{N_1}$ and $Y_L$ with $z$ for $K=100$ and initial condition $Y_{N_1}^{in}=Y_{N_1}^{eq}$ in non-standard leptogenesis and its comparison with standard leptogenesis.}
\label{2}
\end{figure}
In Fig.~\ref{2}(a)-(b), we show the evolution of $Y_{N_1}$ and $Y_L$ with $z=M_1/T$ for two different values of $n=2,4$ and compare the results with standard leptogenesis (SL) in absence of scalar field $\varphi$. We have kept other parameters fixed with values $T_r/M_1=0.01,~K=100$ and $\varepsilon=10^{-6}$ 
and considered RHN has equilibrium number density at $z=0$. We solve the modified Boltzmann equations to obtain the 
abundances $Y_{N_1},~Y_L$ and compare the results with the solutions from ordinary leptogenesis. From Fig.~\ref{2}(a), we observe that co-moving number density $Y_{N_1}$ deviates from the standard leptogenesis case for larger values of $n$. 
For larger values of $n=2$ or $n=4$, due to faster expansion of the Universe $Y_{N_1}$ deviates more from the equilibrium. This makes out of equilibrium decay of RHN more prominent allowing successful leptogenesis. In Fig.~\ref{2}(b), we plot the variation of lepton asymmetry for the same set of parameters. Black horizontal lines indicate the value of $Y_L$ required to generate the Baryon asymmetry in the Universe as obtained by Planck \cite{Patrignani:2016xqp,Ade:2015xua}. It is shown in Fig.~\ref{2}(b) that for normal leptogenesis, the washout of asymmetry is very effective. As a result, the lepton asymmetry fails to produce the observed baryon asymmetry in the Universe. However, the situation changes for $n=2$ and washout effect is reduced since decay parameter is modified. On the other hand for $n=4$, we observe that the washout effect is completely negligible as $K_{\rm eff}<1$ even for larger values of $z\sim 10$ as shown in Fig.~\ref{1}(a). Therefore, a considerable enhancement in lepton asymmetry $Y_L$ can be achieved in modified cosmology depending on the value of $n$.
\begin{figure}[!ht]
\centering
\subfigure[]{
\includegraphics[height=6 cm, width=7 cm,angle=0]{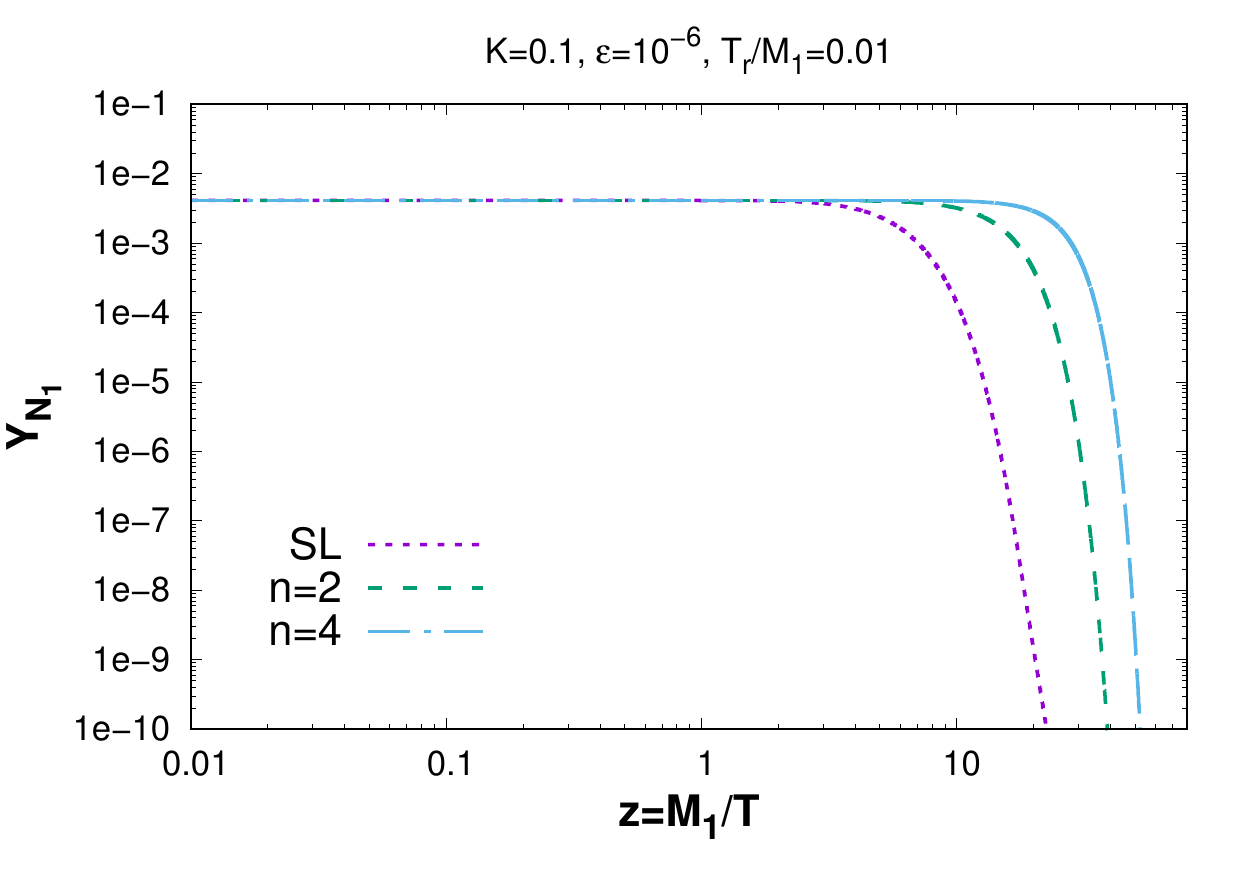}}
\subfigure []{
\includegraphics[height=6 cm, width=7 cm,angle=0]{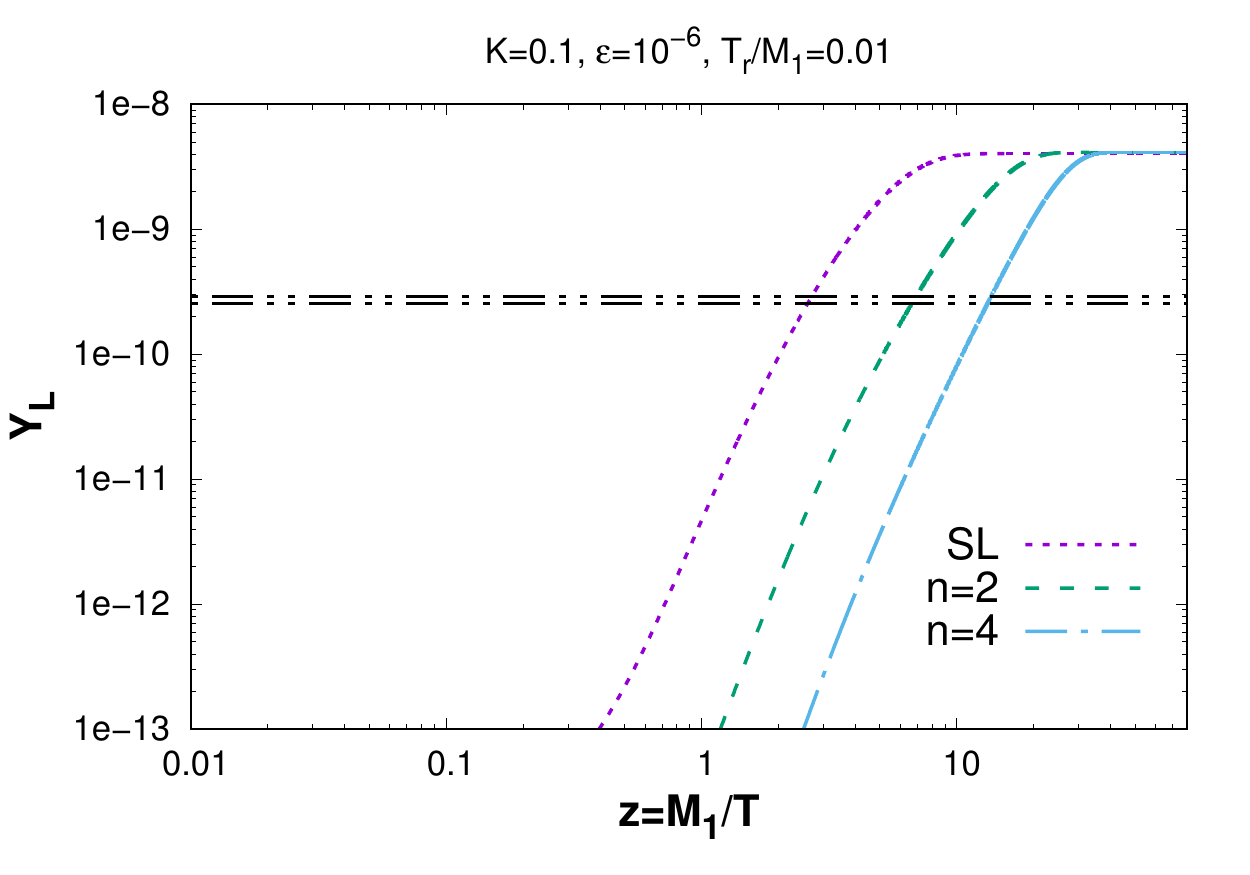}}
\caption{Variation of $Y_{N_1}$ and $Y_L$ with $z$ for $K=0.1$ and $Y_{N_1}^{in}=Y_{N_1}^{eq}$ in non-standard leptogenesis and its comparison with standard leptogenesis.}
\label{3}
\end{figure}
Similar plots for $Y_{N_1}$ and $Y_L$ are shown in Fig.~\ref{3} with $K=0.1$  for same values of $n$ keeping all other parameters fixed. From Fig.~\ref{3} we observe that although there is deviation in evolution of co-moving density of $N_1$ from standard leptogenesis scenario when compared with $n=2$ or $n=4$ case, the yield of lepton asymmetry do not suffer much changes and are same for all three cases. In modified leptogenesis, with increase in $n$, the decay of RHN is delayed due to fast expansion as a result the production of lepton asymmetry also happens at higher values of $z$ as observed in Fig.\ref{3}. Therefore, for $K=0.1$, when leptogenesis is dominated by weak washout, it appears that leptogenesis is almost independent of the modified cosmology. However, this is not always valid and might differ if initial condition is changed, which will be shown in Sec.~\ref{y0}. 
\subsubsection*{Effects of $T_r/M_1$ in BE solutions}
\begin{figure}[!ht]
\centering
\subfigure[]{
\includegraphics[height=6 cm, width=7 cm,angle=0]{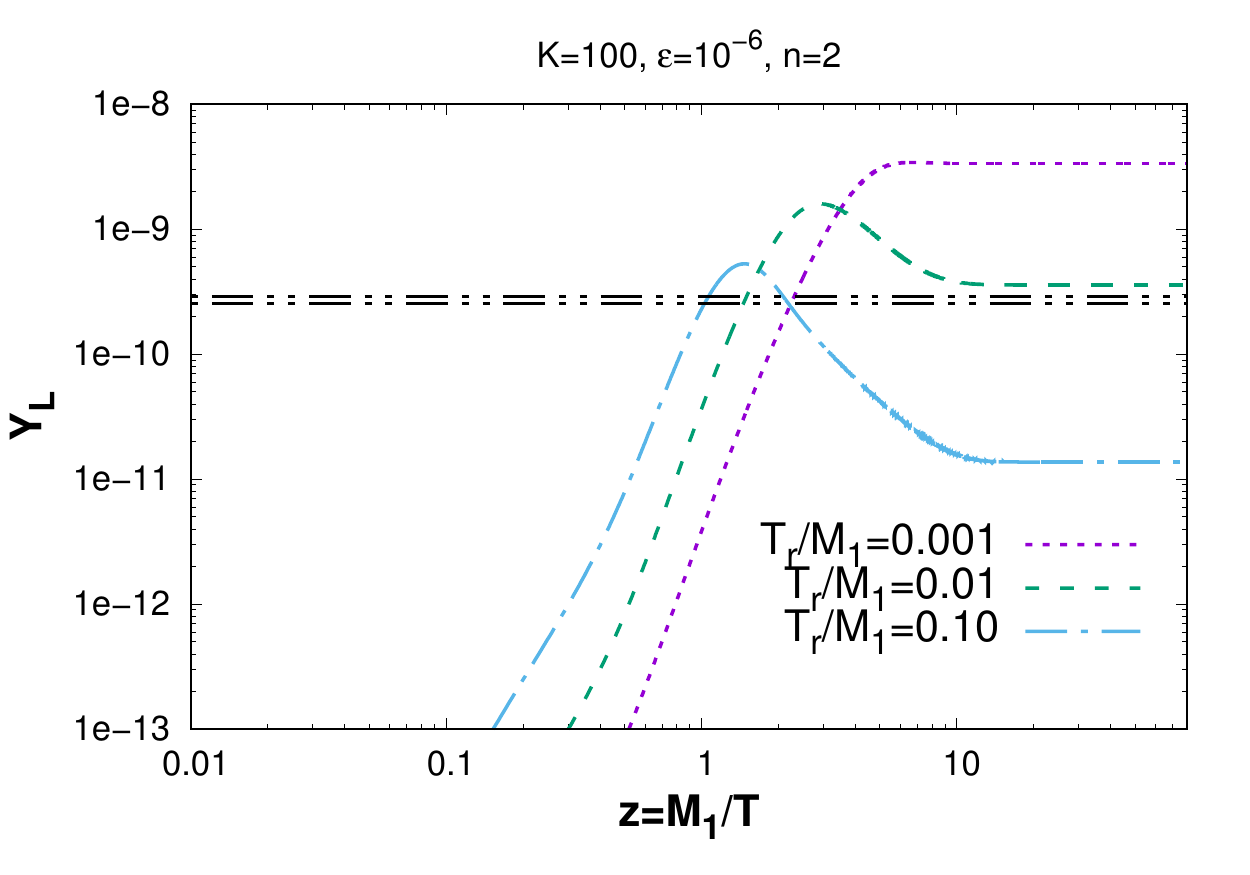}}
\subfigure []{
\includegraphics[height=6 cm, width=7 cm,angle=0]{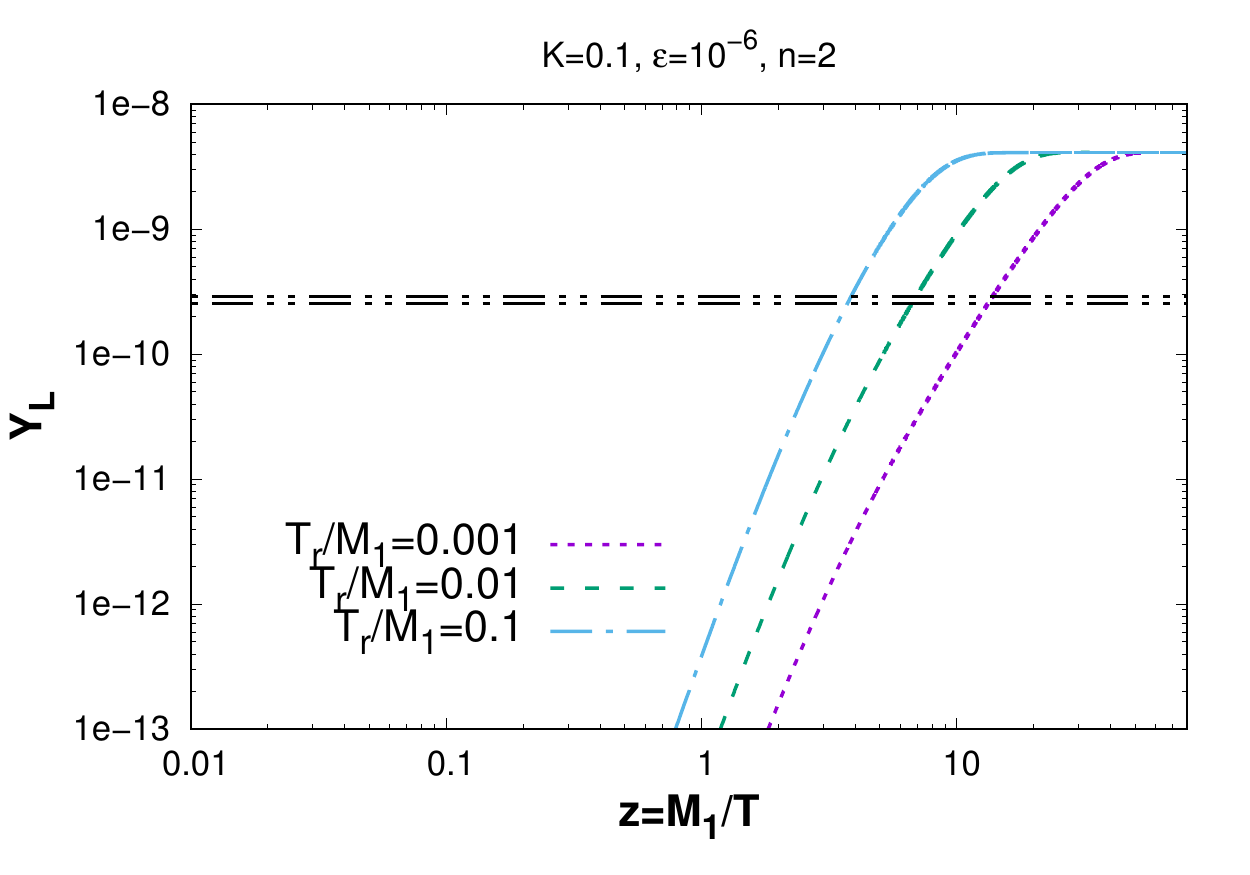}}
\caption{Left panel: $Y_L$ vs $z$ for different $T_r/M_1$ with $K=100$ and $n=2$. Right panel:$Y_L$ vs $z$ for different $T_r/M_1$ with $K=0.1$ and $n=2$.}
\label{4}
\end{figure}
So far we have explored the effects of $n$ on the Boltzmann equations for leptogenesis for some fixed values of other input parameters $T_r/M_1,~K,~\varepsilon$ with the initial condition $Y_{N_1}=Y_{N_1}^{eq}$. Now, we repeat the same study for three different set of $T_r/M_1=0.1,0.01,0.001$ for $n=2$ and $\varepsilon=10^{-6}$. Considering two values of $K=100,~0.1$ we plot in Fig.~\ref{4} the evolution of lepton asymmetry in the present framework of non-standard cosmology. From Fig.~\ref{4}(a), we observe that Yield of co-moving lepton asymmetry
$Y_L$ for $K=100$, suffers a large washout for $T_r/M_1=0.1$ starting near $z\sim 1$ and final asymmetry remains below the required lepton asymmetry to produce the required baryon abundance. However as the ratio decreases by an order ($T_r/M_1=0.01$) the washout effect becomes mild and disappears for $T_r/M_1=0.001$ resulting $Y_L$ value higher than observed baryon abundance. Comparing the lepton asymmetry results in Fig.~\ref{4}(a) with the standard lepton asymmetry results from Fig.~\ref{2}(b), we observe that for a fixed $n=2$, as $T_r/M_1$ increases, the washout effect increases and $Y_L$ is almost similar to standard leptogenesis
for $T_r/M_1=0.1$. Therefore, we can conclude that although larger values of $n$ can suppress the washout effect, it can become prominent if $T_r\sim M_1$.
However, for $K=0.1$, we do not observe any effects on lepton asymmetry with changes in $T_r/M_1$ value. Therefore,  we conclude that, for $K=0.1$ modified Boltzmann equations do not alter the leptogenesis at all while for large $K$, decrease in $T_r/M_1$ reduce the washout of asymmetry and $Y_L$ is enhanced with respect to standard leptogenesis.

\subsection{Case II : $Y_{N_1}^{in}=0$, $\varepsilon=10^{-5}$}
\label{y0}
In the earlier  section we have presented our observations on how modification of BEs for leptogenesis changes the abundance of RHN and lepton asymmetry for different parameters with initial condition $Y_{N_1}^{in}=Y_{N_1}^{eq}$. In this section, we repeat the same analysis for the case with zero initial abundance of RHN ($Y_{N_1}^{in}=0$). Similar to Sec.~\ref{yeq}, we present our results with variation of new parameters $n$ and $T_r/M_1$ keeping fixed CP asymmetry $\varepsilon=10^{-5}$.
\subsubsection*{Effects of $n$ on BE solutions}
\begin{figure}[!ht]
\centering
\subfigure[]{
\includegraphics[height=6 cm, width=7 cm,angle=0]{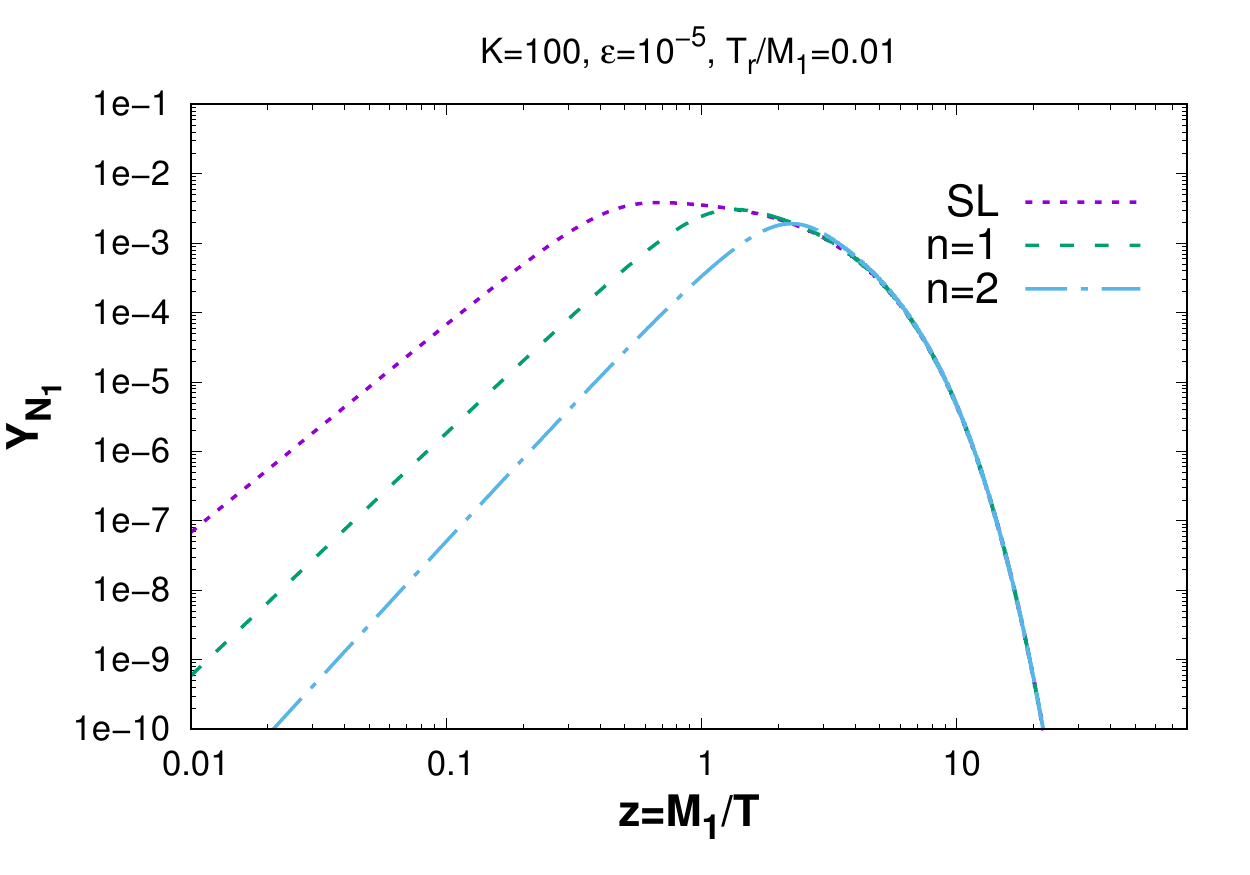}}
\subfigure []{
\includegraphics[height=6 cm, width=7 cm,angle=0]{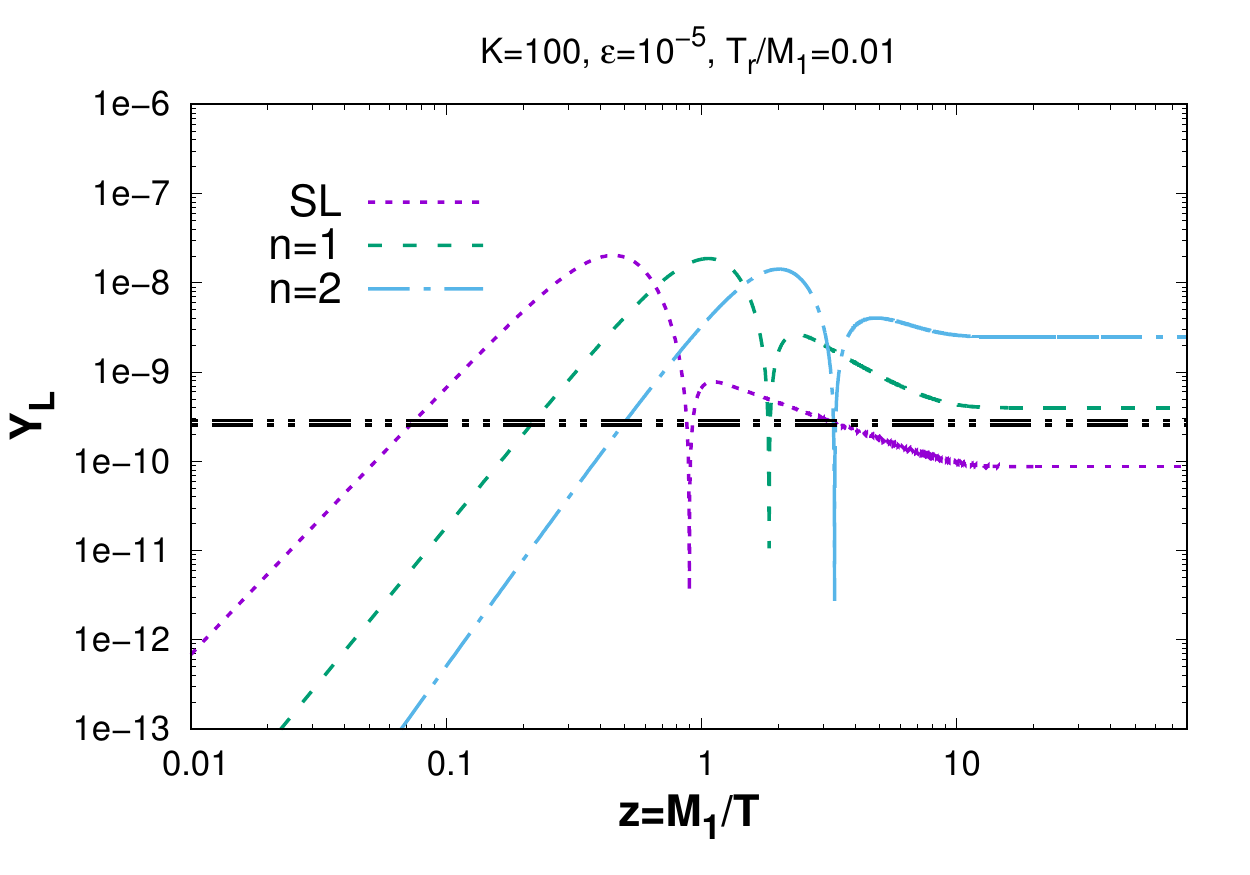}}
\caption{Left panel: Comparison of $Y_{N_1}$ in non-standard leptogenesis with normal leptogenesis plotted against $z$ for $Y_{N_1}^{in}=0$ and $K=100$. Right panel: Similar plots for $Y_L$ against $z$ for same set of parameters obtained for $Y_{N_1}^{in}=0$ and $K=100$.}
\label{5}
\end{figure}
In Fig.~\ref{5}, we show the evolution of comoving abundances of $N_1$ and $Y_L$ with $z$ for vanishing RHN initial abundance. It is interesting to notice from Fig.~\ref{5}(a), that initially RHN abundance increases due to production from inverse decays for small $z<1$ and then starts decaying for $z>1$. As a result a negative asymmetry in $Y_L$ is generated for $z<1$ which is washed out completely after RHN decay takes place for higher values of $z$ resulting a net positive asymmetry. This is usually the case for standard leptogeneis, which is also followed by the modified leptogenesis framework for $n=1$ and $n=2$ case as depicted in Fig.~\ref{5} for larger values of $z$.
We use CP asymmetry parameter $\varepsilon=10^{-5}$ with $T_r/M_1=0.01$ and $K=100$ to generate our results for $Y_{N_1}$
and $Y_L$. From Fig.~\ref{5}(a), we observe that for $z<1$, with increasing $n$, the amount of RHN production is reduced. This is due to the fact that for increasing $n$, washout out from inverse decay gets reduced due to faster expansion. However, for $z>1$ the plots coincides with the standard leptogeneis. 
Also from Fig.~\ref{5}(b), we observe that although the plots with $n=1,2$ follows the similar nature with that of leptogenesis, as the washout is less, the net amount of lepton asymmetry produced increases and $Y_L$ is  order higher for $n=2$ case. 

\begin{figure}[!ht]
\centering
\subfigure[]{
\includegraphics[height=6 cm, width=7 cm,angle=0]{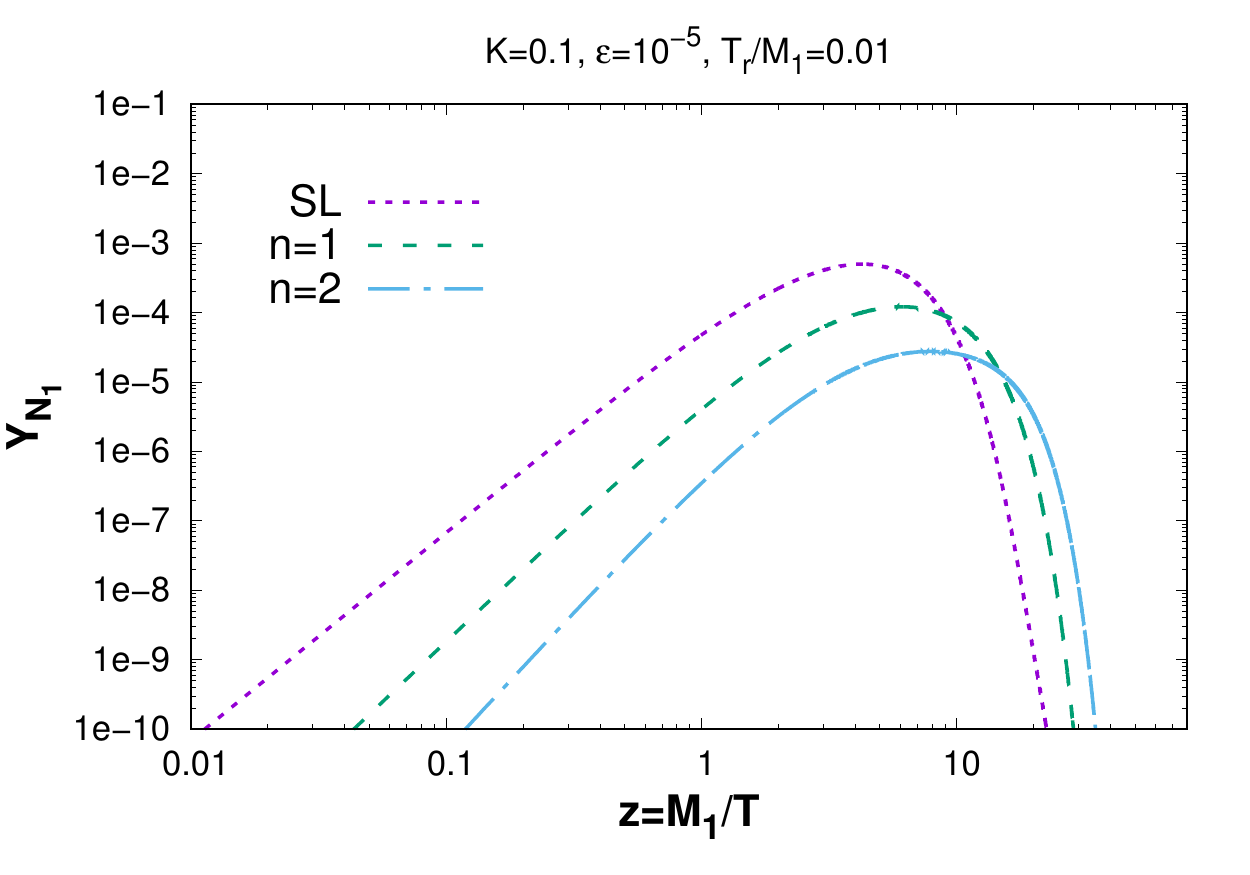}}
\subfigure []{
\includegraphics[height=6 cm, width=7 cm,angle=0]{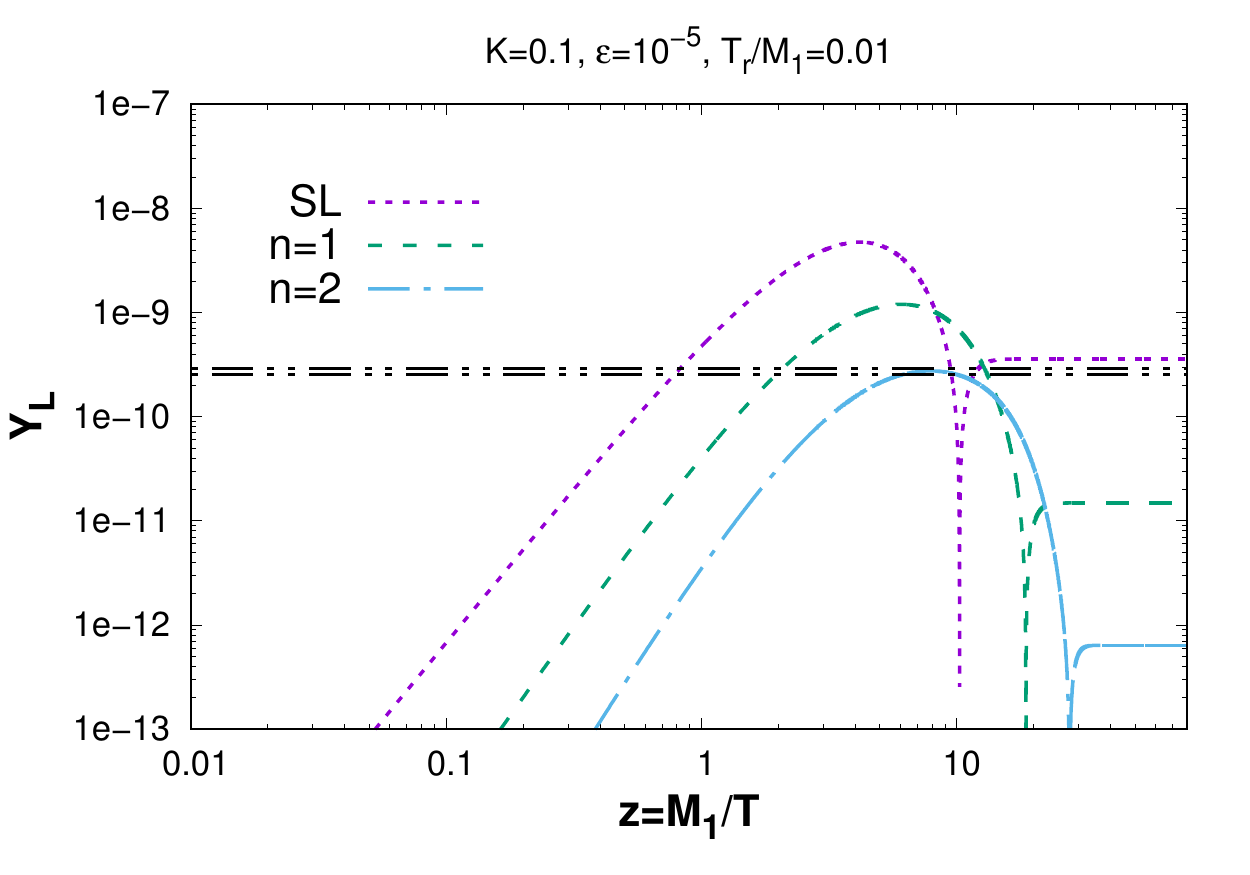}}
\caption{Left Panel: Same as Fig.~\ref{5}a with $K=0.1$ and compared with standard leptogenesis. Right panel: Same as Fig.~\ref{5}b with $K=0.1$.}
\label{6}
\end{figure}

In Fig.~\ref{6}, we repeat the study for $Y_{N_1}^{in}=0$ case with $K=0.1$ keeping all other parameters same as considered for Fig.~\ref{5}. We observe that although the plots for $Y_{N_1}$ in case of $n=1,2$ follows the similar nature of ordinary leptogenesis, the maximum value of $Y_{N_1}$ reached reduces with increasing $n$. From Fig.~\ref{6}(b), we notice that amount of final lepton asymmetry obtained also decreases with higher $n$ value and fails to generate the required abundance of $Y_L$ to explain baryon asymmetry in the Universe. The net lepton asymmetry produced depends on abundance $Y_{N_1}$. The amount of $Y_{N_1}$ is decreased with larger $n$ which also reduces $Y_L$
considerably and $Y_L$ is almost two order of magnitudes smaller when compared with standard leptogenesis.  This is completely opposite to the conclusion from the case with $K=100$ (Fig.~\ref{5}(b)) where $Y_L$ is enhanced with increase in $n$. Therefore, we conclude that
depending on the the decay parameter $K$, for vanishing RHN abundance, the final lepton asymmetry $Y_L$ can be enhance or reduced from standard scenario for a certain value of $n$
and $T_r/M_1$.

\subsubsection*{Effects of $T_r/M_1$ in BE solutions}

\begin{figure}[!ht]
\centering
\subfigure[]{
\includegraphics[height=6 cm, width=7 cm,angle=0]{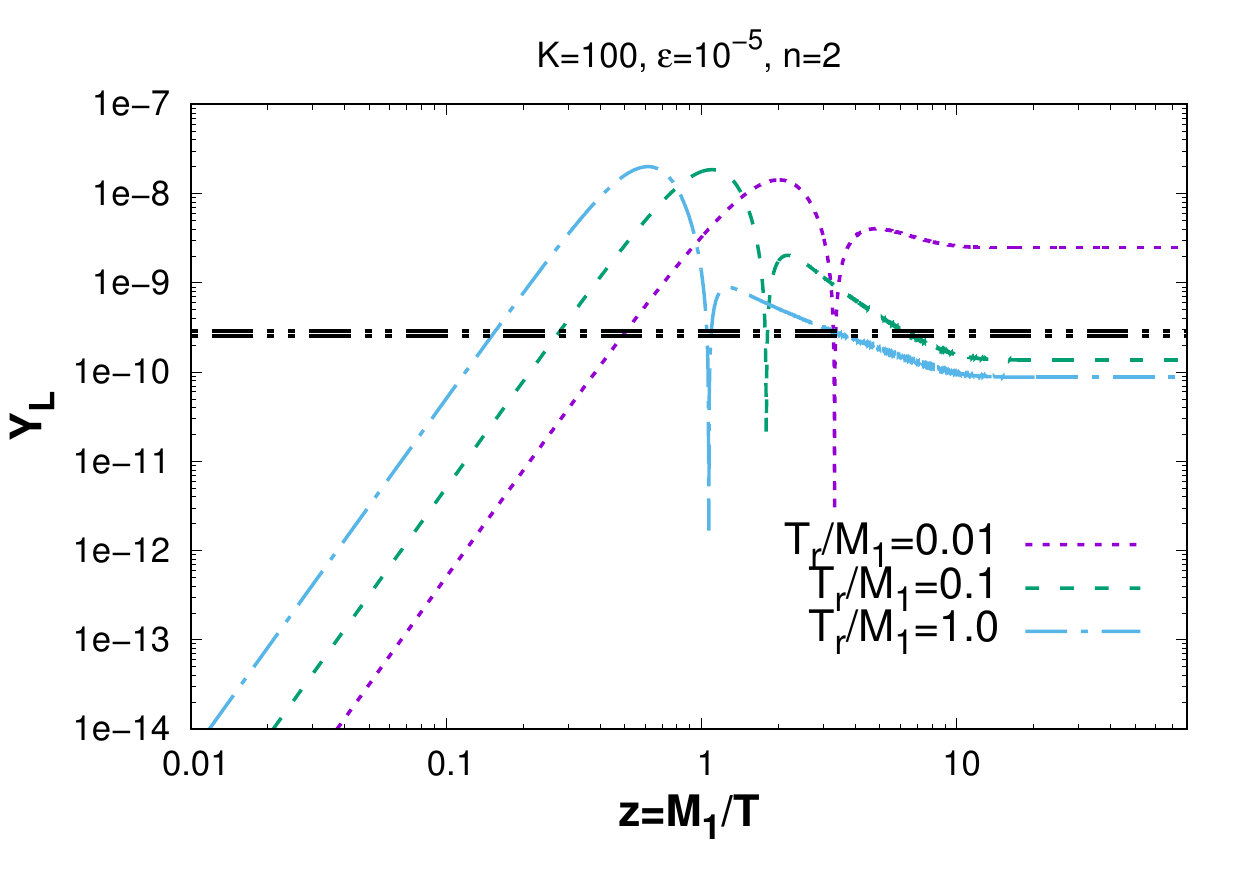}}
\subfigure []{
\includegraphics[height=6 cm, width=7 cm,angle=0]{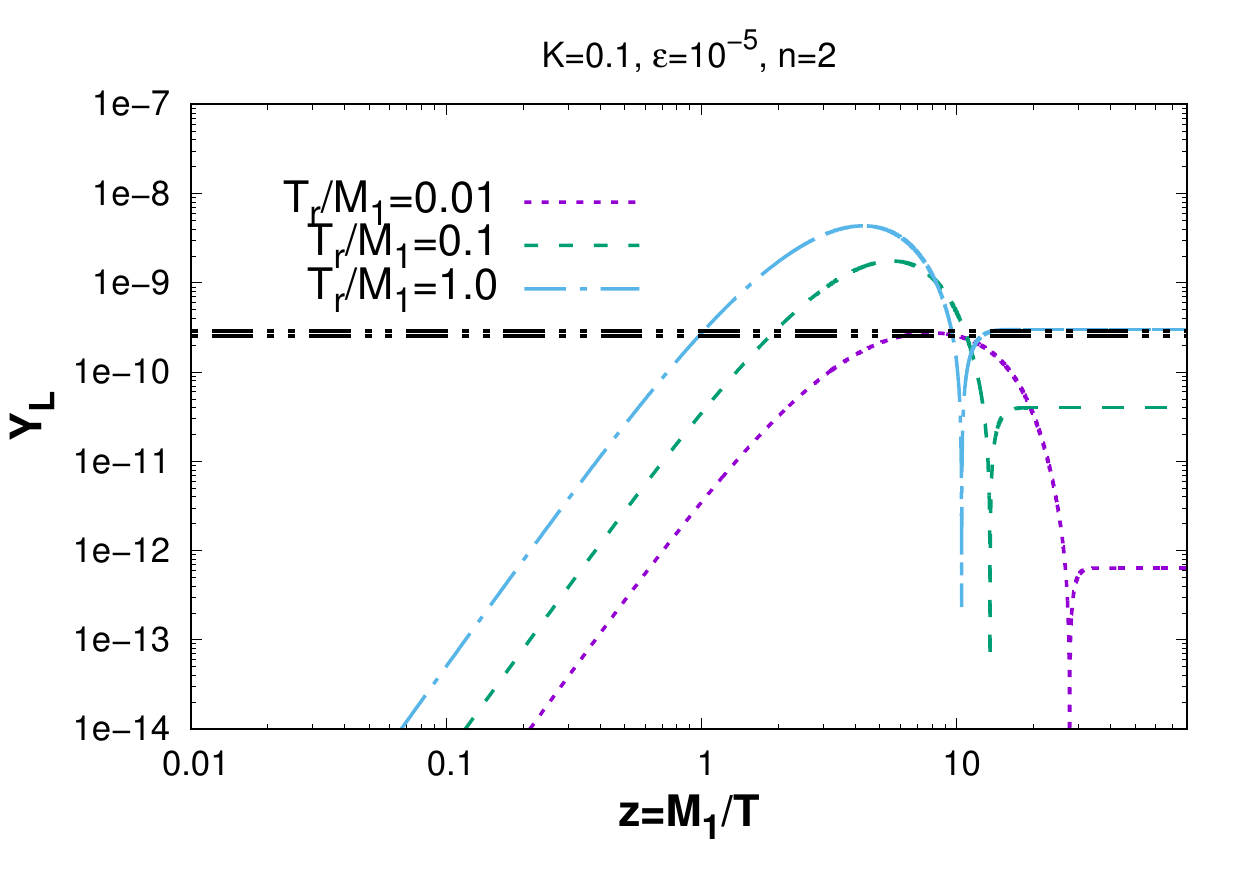}}
\caption{Left panel shows the variation of $Y_{L}$ with $z$ for different $T_r/M_1$ using $Y_{N_1}^{in}=0$ and $K=100$. Similar plot by using $K=0.1$ is plotted in right panel. For both the plots $n$ and $\varepsilon$ values are kept fixed.}
\label{7}
\end{figure}

In Fig.~\ref{7}, we show the variation of lepton asymmetry $Y_L$ with $z$ for three chosen $T_r/M_1=0.01,0.1,1$ keeping other parameters fixed at values $n=2$ and $\varepsilon=10^{-5}$.
From Fig.~\ref{7}(a) it can be concluded that with increasing value of $T_r/M_1$, the lepton asymmetry $Y_L$ reduces for a fixed $n$. In fact comparing with Fig.~\ref{5}(b), it can be noticed that for large $T_r/M_1$ the nature of $Y_L$
evolution almost follows the case of ordinary unfalvoured leptogeneis. 
This is quite natural since with since as $T_r\sim M_1$, the Universe becomes radiation dominated. 
Similar conclusion can be made for the case with $K=0.1$ as shown in Fig.~\ref{7}(b) when compared with standard solution shown in Fig.~\ref{6}(b). However, in this case the final lepton asymmetry enhances with increasing value of $T_r/M_1$. In fact, $Y_L$ is almost two order higher for $T_r/M_1=1$ than the case with $T_r/M_1=0.01$ and satisfies the required lepton asymmetry limit.   

To conclude this section, the standard leptogenesis with unflavored approximation can be significantly changed in presence of the additional scalar $\varphi$ in the era of leptogenesis. 
Considerable enhancement or reduction to the lepton asymmetry $Y_L$ can be achieved depending on new parameters $n,T_r/M_1$ which modify the Boltzmann equations for leptogenesis.

\section{Conclusions}
\label{con}

In this work we have performed a study of leptogenesis in modified cosmology and compared the results with standard leptogenesis. The standard evolution of the Universe is modified
by including a scalar field $\varphi$ which dominates over radiation energy density resulting in fast expansion of the Universe. If the effects of the scalar field is active in the era of 
leptogenesis, it will significantly change the dynamics of lepton asymmetry in the Universe. We solve for the modified Boltzmann equations for leptogenesis to obtain the abundance 
of right-handed neutrinos and lepton asymmetry. We observed that presence of the scalar field introduces new parameters that can control the abundance of right-handed neutrino 
and lepton asymmetry and sufficient enhancement or reduction to lepton asymmetry can occur. One important outcome and distinctive feature of the above study is the deviation in 
amount of lepton asymmetry $Y_L$ generated when compared with standard leptogenesis except for the case with equilibrium initial RHN abundance ($Y_{N_1}^{in}=Y_{N_1}^{eq}$) 
and small value of decay parameter $K\sim0.1$. By sufficient enhancement or reduction 
of $Y_L$ from usual unflavored leptogenesis, fast expansion of the Universe will significantly 
change the allowed range of parameters obtained from different model based analysis 
consistent with neutrino oscillation data. In present work, we have considered leptogenesis 
from right-handed Majorana neutrinos only, which also generate neutrino mass through
 type-I seesaw mechanism. However, the effect of fast expanding Universe discussed in 
 the work is equally applicable to leptogenesis from other scenarios like triplet scalar and 
 triplet fermion models since its independent of the seesaw mechanism of neutrino mass 
 and depends only on Boltzmann equations. The above treatment is also applicable to TeV scale 
 leptogenesis from radiative seesaw models~\cite{Hugle:2018qbw,Mahanta:2019gfe} and 
 extensions of type-I seesaw~\cite{Alanne:2018brf}.

In the present study, the flavor effects on Boltzmann equations are not taken into account and we have presented results for unflavored case only. 
For this purpose, we conservatively set lightest RHN mass to be heavier $M_1 \geq 5\times10^{11}$ GeV.  
The flavor effects takes place when charged lepton Yukawa interactions are in equilibrium. However, in fast expanding Universe flavor effects may be delayed but still effective at some lower temperature. 
This actually validates the limit of RHN mass taken into consideration. Inclusion of the flavor effects along with the non-standard cosmology may provide new interesting results. 
Therefore, detailed study of flavor effects in the aspect of non-standard cosmology is required which is not addressed in the present study.
It is also worth mentioning that, the present study can also be extended for a case of common origin of leptogenesis and dark matter where lepton asymmetry and asymmetry 
in dark sector are generated simultaneously from the decay of heavy right-handed Majorana neutrinos~\cite{An:2009vq,Falkowski:2011xh}.
Study of these scenarios in the framework of modified cosmology may also enlighten some interesting features to be explored in detail.

\vskip 5 mm 

\noindent {\bf Acknowledgments} : 
ADB thanks Anirban Biswas, Arunansu Sil and Rishav Roshan for useful discussions. This work is supported in part by the National Science Foundation of China (11775093, 11422545, 11947235).

\end{document}